# Sequential versus Manifold Bayesian Optimization under Realistic Experimental Time Constraints

Boris N. Slautin,[1,*] Sergei V. Kalinin[1,2,*]

[1] Department of Materials Science and Engineering, University of Tennessee, Knoxville, TN 37923, USA
[2] Pacific Northwest National Laboratory, Richland, WA 99354, USA

**Abstract**

Bayesian optimization (BO) is widely used for autonomous materials discovery, yet its classical sequential formulation is insufficient for design of experimental workflows that often combine parallel or batch synthesis with inherently serial characterization. Methods such as combinatorial spread libraries and printed libraries sample a defined low-D manifold in the chemical space of the system. Here, we introduce a time-aware framework for comparing sequential and manifold BO under experimentally realistic constraints. By explicitly modeling synthesis and characterization times, we define an effective experimental time metric that enables fair, time-normalized benchmarking of optimization strategies. Using numerical experiments in ternary and quaternary compositional spaces, we show that sequential BO remains optimal for short-term experiments or when batching provides no effective time advantage, whereas manifold BO becomes favorable once multiplexed synthesis enables faster accumulation of measurements. We identify a small set of physically interpretable parameters that govern the transition between these regimes. These results establish a general, experimentally grounded framework for selecting optimization strategies in self-driving laboratories and autonomous materials discovery workflows. The accompanying analysis code is publicly available at https://github.com/Slautin/2025_GP_BO_Manifolds.

[*] Authors to whom correspondence should be addressed: bslauti1@utk.edu and sergei2@utk.edu

## I. Introduction

The demand for accelerating materials discovery and its translation into industrial technologies has become a worldwide trend, attracting substantial investments across the globe.[1, 2] This push is tightly coupled to the development of high-throughput (HT) synthesis and rapid, efficient characterization techniques, which enable exploration of vast compositional and processing spaces inaccessible through conventional manual approaches.[3-9]

Among the high throughput methods, a special place belongs to the combinatorial spread library approach.[5, 10-16] In these systems, chemical composition is encoded via spatial position across the library. This approach was introduced as early as the 1960s, however, its practical utility remained limited for decades by characterization bottleneck and data-handling constraints.[14, 17] Recent advances in deposition methods and pipetting robotics have fundamentally changed this landscape, enabling scalable, reproducible, and widely accessible combinatorial spread or printed (or droplet) libraries.[9, 18-20] Note that while conventional combinatorial libraries represent binary or ternary phase diagrams, the use of multiple sources in the physical deposition methods or multiple endmembers in solution synthesis allows to explore 1D and 2D manifolds within higher-dimensional chemical spaces. The use of the thickness or temperature gradient heaters extends this approach towards exploring the processing or growth effects by extending the dimensionality of sampled chemical spaces.[21-24]

The exploration of materials via combinatorial approach requires building balanced experimental workflows, both at the level of single library exploration via the sequential characterization methods and exploration of higher-D chemical and processing spaces via sequential library by library sampling. At both level, the process is myopic, i.e. the state of the system does not depend on previous measurements. This allows for use of classical Bayesian optimization (BO) based methods.[25-28] In its classical form, at each iteration BO the surrogate model is updated with the latest measurement, and a single next measurement - the most informative or most promising point in the design space – is selected for evaluation. This one-by-one policy has proven extremely effective in domains where experiments are inherently serial. Multiple extensions of BO to batch sampling are available; however, it is generally assumed that each point in the parameter space is available.

However, exploration of high-D chemical spaces via combinatorial libraries introduces the constraint that sampling can be performed only along the chosen low D (and often non-linear) manifold.[10, 14] Within this manifold, characterization remains largely serial: techniques such as scanning probe microscopy, electron microscopy, X-ray diffraction, and Raman spectroscopy typically probe one sample or location at a time.[15, 29-36]

This combination of parallel synthesis and serial characterization introduces a structural mismatch with the assumptions underlying classical sequential BO. While stream-based characterization of combinatorial libraries can reduce overhead associated with sample handling and alignment, the overall discovery rate is governed by how efficiently limited characterization resources are allocated across the library. This motivates the development and evaluation of manifold BO, a framework that aligns the decision-making process with the physical realities of HT materials workflows. Understanding when sequential BO remains optimal, and when manifold BO provides clear advantages, is therefore essential for designing next-generation autonomous discovery systems. Here, we delineate the operational limits of manifold BO and to identify the regimes in which the traditional sequential strategy remains advantageous. By analyzing both approaches under realistic constraints with parallel synthesis but inherently serial characterization, we aim to clarify when it is optimal to maintain a classical BO workflow and when it becomes more effective to transition toward manifold BO for accelerated discovery.

**II. Manifold BO model**

The manifold BO model represents an extension of classical Bayesian optimization to scenarios in which each iteration yields not a single new measurement, but a *batch* of measurements drawn from a manifold of compositions (Figure 1). In this setting, the optimization loop shifts from a strictly one-by-one selection strategy to one in which multiple candidate points are proposed and evaluated. Each combinatorial sample is essentially a low-dimensional manifold (in some case, linear cross-section) of the high-dimensional compositional space. Guided by practical experimental constraints, we restrict our analysis to one-dimensional and two-dimensional manifolds and grid measurements along the manifold. A key hyperparameter of the manifold BO framework is the number of measurements in each batch, $M$, which determines the discretization density along the manifold. In the limiting case $M = 1$, the manifold BO formulation converges to the classical one-by-one Bayesian optimization model.

In manifold BO the acquisition function must be adapted to operate on constrained batches. The classical Expected Improvement (EI) generalizes to the batch Expected Improvement (qEI), which quantifies the expected improvement over the current optimum when evaluating a set of points simultaneously. In this formulation, the gain arises from the best point in the batch. Deeper modifications are required for pure exploration strategies. In this case, identifying regions of high uncertainty cannot rely solely on the posterior variance at the sampled points but it must also account for the kernel length scale, which defines the spatial correlation distance. To capture this effect, we employ a kernel-aware information gain (IG) criterion, which explicitly incorporates the correlation structure imposed by the kernel providing a more accurate assessment of poorly predicted regions. Because manifold evaluation is inherently discrete, it is reasonable to normalize the IG by the number of points (M). This normalization reflects the information gained per measurement and ensures fair comparison between manifolds with different discretization densities.

$$nIG(\chi) = \frac{1}{2M} \log \det(\boldsymbol{I}_M + \sigma^{-2} \boldsymbol{K}_{\chi\chi}) \qquad (1)$$

where $\chi = x_1, x_2, \ldots, x_M$ set of M points on the manifold being evaluated, $\boldsymbol{I}_M$ – identity matrix, $\sigma^2$ – noise variance, $\boldsymbol{K}_{\chi\chi}$ covariance matrix with $[\boldsymbol{K}_{\chi\chi}]_{ij} = k(x_i, x_j)$.

At each iteration, manifold BO selects a set of compositions to be synthesized and characterized simultaneously. In our demonstration, the candidate search space was restricted to linear cross-sections of the compositional simplex (or planar sections in the case of two-dimensional manifolds). However, this choice is not fundamental: the geometry of the manifold can be adapted to experimental constraints and may include nonlinear cross-sections or more complex shapes dictated by synthesis capabilities. It is important to note that the shape of the maniford needs to be calibrated and requires the chemical probing of grown libraries.

Because the number of possible manifolds grows combinatorially with dimensionality, exhaustive evaluation of all candidates at each iteration would be computationally prohibitive.

To maintain tractability, we approximate the candidate space using Monte Carlo sampling, generating a finite subset of candidate manifolds at each iteration for acquisition function evaluation (Figure 1a).

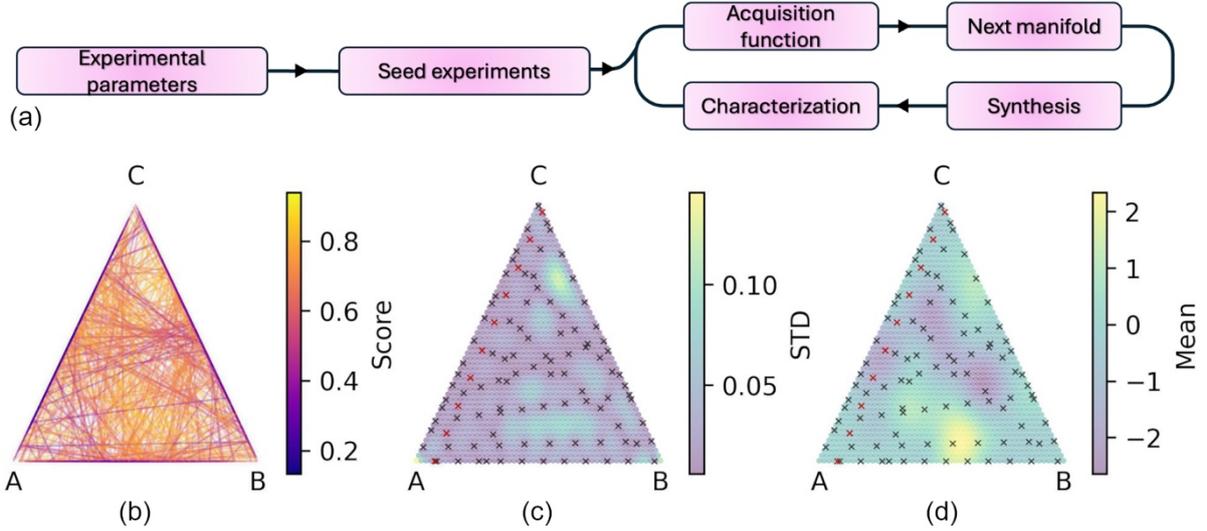

**Figure 1.** (a) Manifold BO framework. (b-d) Exploration of a ternary compositional space. (b) Monte Carlo sampling of candidate manifold cross-sections; the color scale represents the acquisition-function score used to select the next manifold for synthesis and characterization. (c, d) Gaussian process (GP) posterior uncertainty (standard deviation) and posterior mean prediction of the target function across the ternary simplex, respectively. Black crosses indicate compositions that have already been synthesized and measured, while red crosses denote the compositions selected for synthesis and characterization at the next iteration.

### III. Benchmarking Framework

To compare classical point-by-point, *sequential Bayesian optimization* ($BO_S$) with manifold Bayesian optimization ($BO_M$), we conducted a series of controlled numerical experiments using synthetic compositional spaces. As the model system for our combinatorial space, we focus on alloy compositions. As model systems, we employed three-component ($A_xB_yC_{1-x-y}$, 3D simplex) and four-component ($A_xB_yC_zD_{1-x-y-z}$, 4D simplex) alloy design spaces, which represent typical combinatorial composition domains used in high-throughput materials discovery. Each additional element introduces a new compositional degree of freedom, while all fractions remain non-negative and collectively sum to unity. We considered two types of manifolds for batch evaluation: 1D manifolds, representing line-like cross-sections of the compositional simplex, and 2D manifolds, representing planar cross-sections. For simplicity and consistency, all manifolds were constructed such that their endpoints (for 1D) or boundary vertices (for 2D) lay on the edges of the corresponding simplex.

Both optimization strategies employed pure-exploration strategies: the sequential BO used a maximum-uncertainty acquisition function, while the manifold BO used a kernel-aware normalized information gain criterion. Each experiment consisted of 50 independent optimization runs for each strategy intitiated. Each run was initialized with five seed measurements randomly sampled from the compositional spaces and following by the BO-guided exploration. While the total number of acquired points was identical across the compared runs, the efficiency of each strategy was evaluated after normalization to the effective experimental time, described in the following section Operational Model for Experimental Time.

Each run was initialized by generating an effective target function representing a synthetic material property. This function was constructed as a superposition of 106 multivariate Gaussian components, with randomly sampled means within the compositional space and a fixed standard deviation of 0.1. This construction yields a highly heterogeneous and non-trivial landscape, designed to emulate the complexity of realistic composition–property relationships.

**IV. Operational Model for Experimental Time**

Synthesis and characterization proceed on different timescales for the sequential and manifold (high throughput) strategies. To benchmark these strategies and define the practical applicability limits of each approach, we introduce an effective experimental time model. This framework shifts the evaluation of optimization efficiency from iteration-based metrics to time-normalized benchmarking, providing a measure of the true cost of an automated experiment and allowing to optimize sampling strategies.

We define the experimental time required for the synthesis of a single sample in the sequential strategy as $T_1$, and the corresponding characterization time as $T_2$. The total time required to synthesize and characterize one sample, that corresponds to a single iteration of sequential BO$_S$, is:

$$T_S = T_1 + T_2. \qquad (2)$$

The key advantage of the high-throughput synthesis approach is that the time required to fabricate a combinatorial sample is often comparable to that required to produce a single sample in the sequential strategy. We introduce a coefficient $\alpha$ to account for this effect and define the synthesis time for a batch of samples (a manifold) in manifold Bayesian optimization as $\alpha T_1$, where $T_1$ is the synthesis time for a single sample in the sequential case.

While characterization in BO$_M$ remains inherently sequential, practically significant speed-up effects arise due to reduced time for sample transfer, handling, alignment, and instrument setup. We capture this effect by introducing a characterization speed-up coefficient $\beta$, and express the total characterization time for a batch of $M$ compositions as $\beta M T_2$. The total experimental time required to synthesize and characterize a batch of $M$ compositions in the manifold BO strategy is therefore given by

$$T_M = \alpha T_1 + \beta M T_2. \qquad (3)$$

The speed-up factor (S) can be introduced as a ratio between the single iterations in BO$_S$ and BO$_M$, or ratio between time required for the synthesis and characterization of the single sample in sequential and HT cases.

$$S = \frac{T_M}{T_S} = \frac{\alpha T_1 + \beta M T_2}{T_1 + T_2} = \frac{\alpha + \beta M \frac{T_2}{T_1}}{1 + \frac{T_2}{T_1}} \qquad (4)$$

The speed-up factor is determined by the parameters $\alpha$, $\beta$, $M$, and the ratio $T_2/T_1$. For simplicity, we assume $\alpha = 1$, corresponding to the case in which the time required to synthesize a single sample in the sequential strategy is equal to the time required to synthesize a single high-throughput combinatorial sample in the manifold strategy. Practically, the values of these constants can be readily estimated by domain expert before launching the experimental campaign, and are trivially determined based on the experimental traces after the experiment.

To enable comparison between manifold strategies with different numbers of points per manifold, we further normalize the speed-up factor by the number of compositions in the manifold. This normalized speed-up, $S_{\text{norm}} = S/M$, has a clear physical interpretation: it represents the ratio of experimental time required to characterize a single composition in the manifold-based strategy relative to the sequential baseline, rather than the time required to process an entire sample. Using the expressions for $T_S$ and $T_M$, the normalized speed-up factor can be written explicitly as

$$S_{norm} = \frac{S}{M} \approx \frac{1+\beta M \frac{T_2}{T_1}}{M(1+\frac{T_2}{T_1})}. \tag{5}$$

In the limiting case $\beta = 1$ and $M = 1$, manifold BO reduces to the sequential strategy, yielding $S_{\text{norm}} = 1$. For a fixed relative characterization dominance $T_2/T_1$, the normalized speed-up $S_{\text{norm}}$ increases linearly with $\beta$ (Figure 2a), reflecting the direct impact of characterization overhead on per-composition efficiency.

Conversely, for fixed $\beta$, three distinct regimes can be identified as a function of $T_2/T_1$ (Figure 2b). In the synthesis-limited regime ($T_2/T_1 \ll 1$), the normalized speed-up approaches

$$S_{\text{norm}} \approx \frac{1}{M} \text{ (or } \frac{\alpha}{M} \text{ if } \alpha \neq 1\text{).} \tag{6}$$

In this regime, the per-composition cost is determined by the synthesis step, making batching largely insensitive to the characterization speed-up factor $\beta$. In the opposite, characterization-limited regime ($T_2/T_1 \gg 1$), the normalized speed-up approaches

$$S_{\text{norm}} \approx \beta. \tag{7}$$

In this limit, the achievable acceleration is governed primarily by the characterization stage, and improvements in synthesis throughput provide little additional benefit. In the intermediate regime ($T_2/T_1 \sim 1$), the normalized speed-up is well approximated by

$$S_{\text{norm}} \approx \frac{1+\beta M}{2M}, \tag{8}$$

indicating a smooth transition between synthesis-dominated and characterization-dominated limits. The resulting landscape of the normalized speed-up $S_{\text{norm}}$ as a function of $T_2/T_1$ and $\beta$ is shown in Figure 2c.

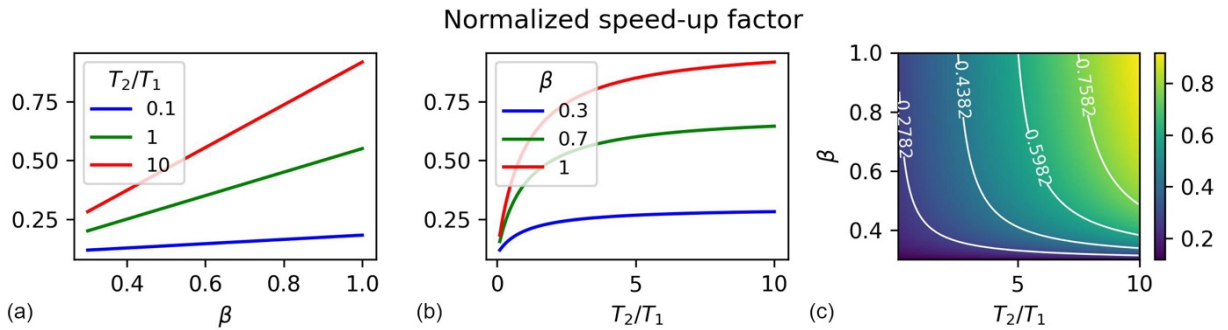

**Figure 2.** Normalized speed-up factor. (a) $S_{\text{norm}}$ versus $\beta$ at fixed $T_2/T_1$. (b) $S_{\text{norm}}$ versus $T_2/T_1$ at fixed $\beta$. (c) $S_{\text{norm}}$ in the $T_2/T_1$–$\beta$ parameter space. $M = 10$.

# V. Sequential versus Manifold BO strategies

In the following section, we systematically compare the performance of $BO_S$ with $BO_M$ across three representative scenarios. Firstly, we examine a ternary alloy compositional space sampled through a 1D linear combinatorial library. Then, we extend the study to a quaternary (4D) alloy system. Exploration performance was monitored primarily through the average posterior uncertainty, quantified by the average GP standard deviation (STD) across the compositional space. As a complementary metric, we also evaluated the absolute average prediction error between the GP posterior mean and the ground-truth objective function (Abs Error).

As the discrete time unit, we take the duration of a single iteration in sequential Bayesian optimization and normalize it to unity, $T_S = 1$. Under this convention, the cumulative experimental time after $N^{(S)}$ iterations of sequential BO is $T_S^{(N)} = N^{(S)}$. For manifold BO, the corresponding cumulative experimental time after $N^{(M)}$ iterations, where each iteration corresponding to the characterization of one combinatorial sample, is scaled by the speed-up factor $S$ and can be expressed as $T_M^{(N)} = S\,N^{(M)}$. Importantly, while $N^{(S)}$ directly corresponds to the number of characterized compositions in the sequential strategy, each iteration of manifold BO yields $M$ characterized compositions. Therefore, the total number of characterized compositions acquired by $BO_M$ after $N^{(M)}$ iterations is given by $N_{\text{comp}}^{(M)} = M\,N^{(M)} = \frac{M}{S} T_M^{(N)}$.

We first performed a detailed comparison of manifold and sequential exploration strategies within the compositional space of a ternary alloy. For the manifold BO strategy, we set the number of points per manifold to $M = 15$. Each optimization run for both strategies consisted of 150 measured compositions in total. To account for differences in experimental throughput, the results were normalized using the normalized time model, with the normalized speed-up factor $S_{\text{norm}}$ varied from 0.3 to 1. Comparisons were restricted to the time interval over which all strategies yielded available data points, since the effective duration of the $BO_M$ runs decreases with decreasing $S_{\text{norm}}$.

A decrease in $S_{\text{norm}}$ reflects an increasing acceleration of the synthesis-characterization cycle in the manifold strategy relative to the sequential policy. This acceleration leads to a more rapid accumulation of measured compositions, thereby enhancing the effectiveness of manifold BO. As expected, for $S_{\text{norm}} = 1$, corresponding to equal synthesis and characterization time per composition, the sequential strategy outperforms manifold BO across the entire experimental time range in terms of both posterior uncertainty (STD) and absolute prediction error (Figure 3a,b). In the absence of any time advantage, sequential point selection enables more informed decision-making, resulting in better performance. In contrast, in the opposite case $S_{\text{norm}} = 0.3$, manifold BO consistently outperforms the sequential strategy (Figure 3a,b). This behavior reflects the substantially larger number of compositions explored by manifold BO at the same effective experimental time, which outweighs the loss of per-step adaptivity inherent to batch selection.

To assess the relative performance of the two strategies as a function of the speed-up factor $S_{norm}$ and experimental time, we analyze the differences in posterior uncertainty (STD) and absolute prediction error between BO$_S$ and BO$_M$. Negative values (shown in blue) correspond to regions where both the STD and absolute error of the sequential strategy are lower than those of manifold BO, indicating an advantage of BO$_S$. Conversely, positive values (shown in red) highlight regions in which manifold BO provides superior performance. To suppress statistical noise, differences with absolute values smaller than 0.02 – corresponding to the lowest 10% of the overall value range – are set to zero. These regions, marked in white, are attributed to parameter regimes in which both strategies exhibit comparable performance within statistical uncertainty.

At the early stages of the experiment, when the number of measured compositions is still small and results of differences in selection strategies have not yet become significant, both strategies exhibit comparable performance in terms of posterior uncertainty reduction across a broad range of $S_{\text{norm}}$. As the experiment progresses, a clear separation between the strategies emerges. For $S_{\text{norm}} \gtrsim 0.6$–$0.7$, sequential BO maintains an advantage in reducing posterior uncertainty throughout the entire exploration (Figure 3c). In contrast, for smaller values of $S_{\text{norm}}$, a transition is observed from a regime of comparable performance or sequential BO advantage toward one in which manifold BO becomes superior (Figure 3c). This transition occurs earlier as $S_{\text{norm}}$ decreases. The transition point reflects the balance between adaptive point selection and sampling density: it marks the stage at which the increased number of measured compositions in manifold BO outweighs the benefit of more informed, sequential composition selection. A similar trend is observed for the absolute-error–based performance metric (Figure 3d). In this case, the transition toward the superiority of manifold BO occurs earlier than in the STD-based analysis. This behavior reflects the greater sensitivity of absolute prediction error to differences in the number of measured compositions, compared to posterior uncertainty.

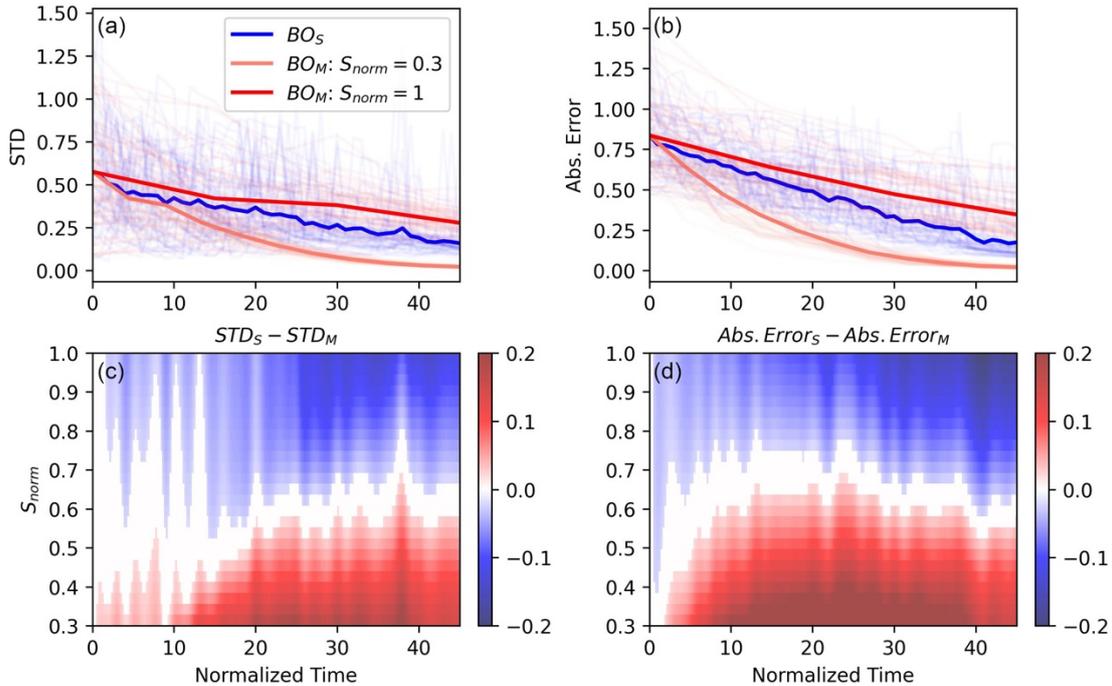

**Figure 3.** Comparison of sequential and manifold BO. Average performance over 50 independent runs for (a) posterior uncertainty (GP standard deviation, STD) and (b) absolute

prediction error as a function of effective experimental time for BO$_S$ and BO$_M$ at $S_{\text{norm}} = 1$ and $S_{\text{norm}} = 0.3$. (c,d) Difference maps for STD and absolute error, respectively. Red regions indicate superior performance of BO$_M$, blue regions indicate superior performance of BO$_S$, and white regions correspond to statistically comparable performance. M = 15.

In the next step, we extended the analysis to investigate the effect of varying the number of measured compositions per manifold, $M$, for higher-dimensional compositional spaces. Linear manifolds were employed for both the ternary and quaternary systems. Performance was assessed using difference maps that highlight regions of superior performance for different parameter regimes (Figure 4). At this stage, we did not directly compare manifold BO strategies with different values of $M$ against each other. Instead, each manifold BO configuration was analyzed relative to the sequential BO baseline. For both 3D and 4D compositional spaces, increasing $M$ systematically shifts the transition point, where manifold BO begins to outperform sequential BO, toward longer effective experimental times. This shift reflects the growing disparity between the two strategies: as $M$ decreases, manifold BO progressively approaches the sequential limit and converges to BO$_S$ for $M = 1$. A similar shift of the transition point toward longer times is observed when moving from 3D to 4D compositional spaces. This trend can be attributed to the increased dimensionality of the design space, which requires a larger number of sampled compositions for adequate characterization.

From a practical perspective, this behavior indicates that manifold BO is most advantageous in long-term exploration scenarios, where a sufficiently large number of compositions must be characterized before the experiment concludes. While the normalized speed-up factor $S_{\text{norm}}$ is the fundamental parameter governing this behavior, we additionally visualize the dependence on the more experimentally intuitive quantities $\beta$ and $T_2/T_1$, with the remaining parameters held fixed. All representations consistently preserve the general trends discussed above.

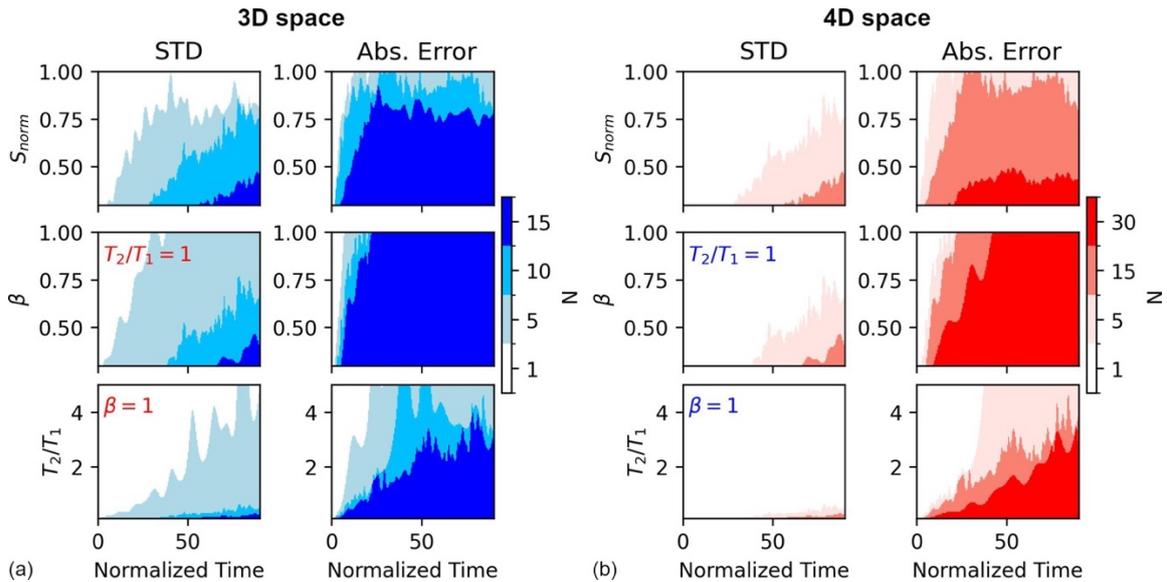

**Figure 4.** Comparative performance maps for BO$_S$ and BO$_M$. Regions indicating favorable performance of sequential and manifold BO for different batch sizes in (a) 3D and (b) 4D compositional spaces, shown as functions of $S_{\text{norm}}$, $\beta$, and $T_2/T_1$ and normalized time.

A direct comparison between manifold BO strategies with different batch sizes indicates that manifolds with smaller $M$ tend to outperform those with larger $M$. However, these results should be interpreted with caution. This apparent advantage is likely a consequence of the simplified experimental time model employed in this study.

In particular, the current model does not explicitly account for the internal structure of characterization delays associated with sample exchange and reconfiguration. In practice, varying the number of compositions per manifold alters the frequency of sample changes, which in turn affects characterization overhead. By treating the characterization speed-up coefficient $\beta$ as identical across different values of $M$, we effectively assume the same per-composition characterization rate regardless of batch size. As a result, comparisons between manifold BO strategies with different $M$ values may reflect modeling assumptions rather than intrinsic algorithmic advantages. Accounting for these effects would necessitate additional parameters describing sample exchange and characterization overhead, leading to a more complex model. Such refinements are more appropriate for case-specific implementations with well-characterized experimental timings, rather than for the general framework presented here. For this reason, we focus primarily on comparisons between manifold BO and the sequential baseline, where the interpretation remains robust.

In the final step of our analysis, we investigated the effect of manifold dimensionality. From a practical standpoint, the most relevant manifold dimensionalities are one- and two-dimensional, as these can be readily realized using spread or gradient-based combinatorial libraries. While higher-dimensional manifolds can, in principle, be constructed in the form of discrete libraries, such implementations are considerably less common in practice. Accordingly, we restrict our analysis to 1D and 2D manifolds, each containing the same number of measured compositions $M = 21$, and compare their performance in exploring a four-dimensional compositional space.

It is evident that the use of two-dimensional manifolds consistently outperforms one-dimensional manifold-based exploration in terms of both posterior uncertainty reduction and absolute prediction error (Figure 5). For both metrics, the transition from sequential BO-dominated regimes to manifold BO-dominated regimes occurs at earlier effective times when 2D manifolds are employed. The two-dimensional manifolds also outperformed sequential BO in reducing absolute error across all values of $S_{\text{norm}}$ over the majority of the experiment.

This behavior indicates that increasing the dimensionality of the explored manifold within a high-dimensional compositional space leads to more favorable sampling distributions, thereby improving the efficiency of combinatorial space exploration. This improvement arises from the reduced redundancy and improved space-filling properties of higher-dimensional manifolds, which yield more informative and less correlated batches of samples in high-dimensional compositional spaces.

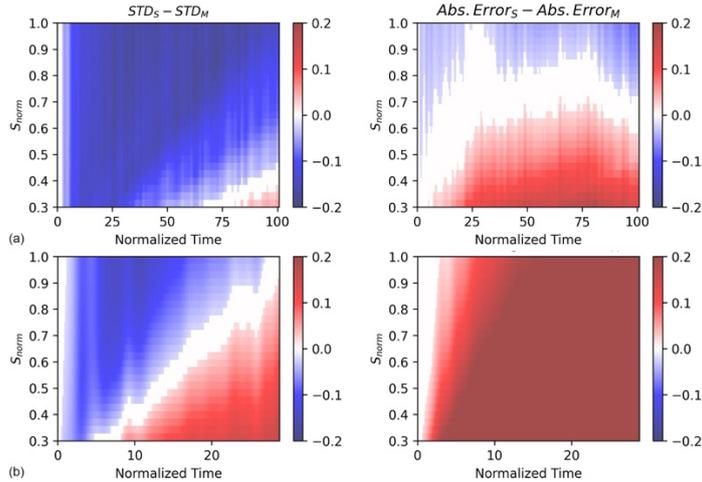

**Figure 5.** Difference maps for STD and absolute error for (a) 1D and (b) 2d manifolds, respectively. Red regions indicate superior performance of $BO_M$, blue regions indicate superior performance of $BO_S$, and white regions correspond to statistically comparable performance. M = 21.

**VI. Scenarios of the practical implementation**

To illustrate the applicability of manifold Bayesian optimization (BO) under realistic experimental conditions, we analyzed several representative closed-loop automated materials discovery scenarios subject to practical time constraints. Specifically, we considered the following characterization techniques as baseline examples: atomic force microscopy (AFM), photoluminescence (PL), laboratory X-ray diffraction (XRD), and synchrotron-based XRD.

Based on characteristic latency estimates for each technique, we evaluated the effective characterization speed-up coefficients $\beta$ for different numbers of measurements within a manifold (Table 1). To perform this estimation, we introduced an additional parameter $T_s$, representing the time required for sample exchange, instrument alignment, and initial tuning prior to measurement.

For most characterization techniques considered, the per-point measurement time $T_2$ is smaller than the overhead time $T_s$. As a result, batching multiple measurements within a single sample significantly amortizes the overhead, leading to a pronounced reduction in effective per-composition characterization time and therefore a strong potential advantage for manifold BO. The primary exception is laboratory XRD in the configuration considered here, where the typical spectrum acquisition time exceeds the estimated sample-change overhead. In this case, the relative benefit of batching is reduced, and the time advantage of manifold BO becomes less pronounced compared to other techniques.

Importantly, all latency values used in this analysis represent order-of-magnitude estimates characteristic of academic research environments. For certain characterization techniques, the measurement time can vary over more than an order of magnitude, depending on material properties, signal strength, spatial resolution requirements, and specific experimental settings. Therefore, the values employed here are intended to capture qualitative regime behavior and relative scaling trends, rather than system-specific performance limits or optimized industrial implementations.

**Table 1.** Estimates for the characterization time parameters for the different characterization techniques

|          | $T_2$, min | $T_s$, min | $\beta$ | | |
|---|---|---|---|---|---|
|          |            |            | N=10 | N=30 | N=100 |
| SPM      | ~5         | ~15        | 0.33 | 0.28 | 0.26 |
| PL       | ~0.1       | ~5         | 0.12 | 0.05 | 0.03 |
| Lab. XRD | ~20        | ~10        | 0.7  | 0.68 | 0.67 |
| Sync. XRD| ~0.02      | ~30        | 0.1  | 0.03 | 0.01 |

On the synthesis side, we consider three representative high-throughput approaches: pulsed laser deposition (PLD) and co-sputtering, both enabling the fabrication of spread combinatorial thin-film libraries, as well as robotic droplet-based synthesis for the creation of discrete composition libraries. For these synthesis platforms, we evaluated the normalized speed-up parameter $S_{\text{norm}}$ for various combinations with the characterization techniques discussed above (Table 2). In these estimates, we assumed a representative batch size of $N = 10$ compositions per manifold.

**Table 2.** Estimates of normalized speed-up coefficients for the combination of the different synthesis and characterization techniques

|               | $T_1$, min | $S_{norm}$, N=10 | | | |
|---|---|---|---|---|---|
|               |            | SPM  | PL  | Lab. XRD | Sync. XRD |
| PLD           | 120        | 0.13 | 0.1 | 0.22     | 0.1       |
| Co-sputtering | 180        | 0.12 | 0.1 | 0.19     | 0.1       |
| Droplets      | 30         | 0.19 | 0.1 | 0.4      | 0.1       |

Based on first-order latency estimates, manifold BO shows a potential time advantage across all considered synthesis–characterization combinations. The strongest advantage (lowest $S_{\text{norm}}$) is observed for PL and synchrotron-based XRD, where the per-measurement acquisition time is substantially smaller than both the synthesis time and the sample-change overhead. In these regimes, batching effectively amortizes the dominant time contributions. A particularly illustrative comparison arises between laboratory X-ray diffraction and synchrotron-based XRD. The characteristic latency profiles for these techniques differ substantially and lead to different optimization regimes under the proposed time model.

For laboratory XRD, the estimated per-point acquisition time is comparable to or even exceeds the sample-change overhead. As a result, the effective characterization speed-up factor $\beta$ remains relatively high, and batching provides only a moderate reduction in per-composition time. Under these conditions, the normalized speed-up parameter $S_{\text{norm}}$ is close to 0.7. Consequently, based on the calculations above, manifold BO becomes advantageous primarily in long-term exploration campaigns, where the cumulative benefit of accelerated data acquisition outweighs the reduced adaptivity inherent to batch selection. For short experimental runs, sequential BO retains an advantage due to its fully adaptive point-by-point decision-making. In contrast, synchrotron-based XRD exhibits extremely short per-measurement acquisition times but significantly larger sample-change and alignment overheads. In this case, batching multiple measurements within a single combinatorial library effectively amortizes the dominant overhead term, resulting in a substantially reduced $\beta$ coefficient and a lower $S_{\text{norm}}$. Therefore, manifold BO becomes particularly attractive for synchrotron campaigns, where rapid accumulation of measurements during limited beamtime is critical.

However, practical accessibility further differentiates these techniques. Laboratory XRD systems are typically readily available and allow flexible, extended experimentation. In such settings, time constraints are less severe, and the adaptive advantages of sequential BO can be fully exploited. By contrast, synchrotron beamtime is highly competitive and limited in duration. Under these constrained conditions, maximizing information gain per unit beamtime becomes paramount and should be also taken into consideration.

**VII. Perspectives**

The rapid emergence of the self-driving labs and lab automation as well as the renaissance of the combinatorial methods necessitates multilevel workflow planning to optimally combine throughputs and costs of the synthesis and characterization steps. The framework here defines a foundation for selecting between sequential and manifold BO under experimentally realistic constraints for exploring high-D compositional spaces via combinatorial spread or printed libraries. The present model represents a deliberately minimal baseline, and several extensions naturally follow.

First, the current formulation assumes grid-like sampling within each selected manifold. In practice, manifold BO can be further strengthened by introducing adaptive intra-manifold sampling strategies. For example, a secondary BO loop could be implemented to adaptively select the most informative compositions for measurement within the chosen manifold. Such hierarchical decision-making would preserve the synthesis-level batching advantage while effectively shifting the characterization stage toward a sequential BO regime without introducing additional time costs.

Second, the latency model can be further refined to account for additional experimental complexities. Instrumental drift, calibration cycles, beamline instability, and signal-to-noise–dependent acquisition times can introduce variations in the measurement latency. Incorporating such time-dependent or stochastic effects would improve the realism of the framework. Moreover, the current formulation assumes ideal knowledge of the compositions. In practice, the actual compositions often deviates from the nominal targets. As a result, realistic autonomous workflows may require intermediate compositional verification steps, which introduce additional latency and may shift the balance between sequential and manifold optimization strategies.

The comparison between laboratory and synchrotron-based XRD illustrates a broader, multi-modal dimension. Synchrotron measurements offer extremely short per-point acquisition times but are constrained by limited beamtime availability and higher operational costs. Laboratory XRD, while slower per measurement, provides accessibility and temporal flexibility. This trade-off naturally defines a Pareto front in the joint space of time and cost. Extending the present model to explicitly incorporate economic constraints would enable optimization strategy selection under budget-limited and beamtime-limited.

More generally, real-world autonomous pipelines rarely consist of a single closed loop. Instead, they involve multi-stage processes: rapid screening using accessible tools, targeted high-resolution measurements, and eventual refinement through high-quality single-composition synthesis. In such hierarchical workflows, manifold BO may dominate during early exploratory stages, where broad coverage is critical, while sequential BO may become preferable during final precision optimization. Embedding the proposed time-aware framework within multi-stage decision architectures represents a natural next step.

Finally, the present analysis is restricted to compositional variables. In practice, materials discovery operates in hybrid spaces that combine composition with processing conditions such as temperature, pressure, atmosphere, deposition rate, and annealing protocols. These conditional parameters effectively expand the search space and often introduce strong coupling between synthesis and structure formation. Manifold BO may become even more advantageous in such extended spaces, where combinatorial synthesis naturally spans multiple correlated variables.

Manifold BO should therefore not be viewed as a universal replacement for sequential strategies, but as a complementary paradigm whose advantage emerges under identifiable physical conditions. The future of autonomous materials discovery will likely involve adaptive combinations of both approaches, selected dynamically in response to evolving experimental constraints.

## VIII. Summary

In conclusion, we introduced a time-aware framework for comparing sequential and manifold Bayesian optimization under experimentally realistic conditions that combine high-throughput synthesis with inherently serial characterization. By explicitly accounting for experimental time costs, we demonstrated that sequential Bayesian optimization remains optimal in the absence of a batching-induced time advantage or in short-term experimental campaigns, whereas manifold strategies become favorable once accelerated synthesis or characterization enable more rapid accumulation of measurements.

We further showed that the transition between these regimes is governed by a small set of physically interpretable parameters, including characterization efficiency, relative synthesis–characterization times, batch size, and manifold dimensionality. In particular, higher-dimensional manifolds yield improved exploration efficiency through reduced redundancy and enhanced space-filling in high-dimensional design spaces.

Together, these results establish a general and practically grounded framework for selecting between sequential and manifold Bayesian optimization strategies. The proposed model can be directly integrated into self-driving laboratory workflows, providing actionable guidance for the design of efficient autonomous materials discovery pipelines under realistic experimental constraints.

**Acknowledgements**

This material (BNS, SVK) is based upon work supported by the National Science Foundation under Award No. NSF 2523284.

**Author Contributions**

**BNS**: Conceptualization; Software; Writing – original draft. **SVK**: Conceptualization; Supervision; Writing – review & editing.

**Data Availability Statement**

The analysis codes and simulation results that support the findings of this study are available at https://github.com/Slautin/2025_GP_BO_Manifolds.